\documentclass[epj]{svjour}

\usepackage{graphicx}
\usepackage{fancyhdr}
\def\makeheadbox{{
 }}
\def\mytitle{My title}
\def\myauthors{My name}
\def\mytype{My type of session}
\def\mysession{My session}

\def\mytitle{Long-living superpartners in the MSSM} 
\def\myauthors{A.V. Gladyshev, D.I. Kazakov and M.G. Paucar}    
\def\mytype{Contributed Talk}
\def\mysession{Colliders - SUSY Phenomenology}

\begin{document}
\title{Long-living superpartners in the MSSM}
\author{A.V. Gladyshev\inst{1,2,}\thanks{\emph{E-mail:} gladysh@theor.jinr.ru},
D.I. Kazakov\inst{1,2}
\and
M.G. Paucar\inst{1}}
\institute{Bogoliubov Laboratory of
Theoretical Physics, Joint Institute for Nuclear Research,\\
141980, 6 Joliot-Curie, Dubna, Moscow Region, Russian Federation
\and
Institute of Theoretical and Experimental Physics,\\
117218, 25 Bolshaya Cheremushkinskaya, Moscow, Russian Federation}
%
\date{}
\abstract{The parameter space of the Constrained Minimal supersymmetric Standard Model is considered.
It is shown that for the particular choice of parameters there are some regions where
long-living charged superparticles exist. Two regions of interest are
the co-annihilation region with light staus, and the region with large negative
trilinear scalar coupling $A$ distinguished by light stops. The phenomenology of
long-living superparticles is briefly discussed.
\PACS{
      {12.60.Jv}{Supersymmetric models}   \and
      {14.80.Ly}{Supersymmetric partners of known particles}
     } 
} 
\maketitle
%
\section{Introduction}
\label{sec:intro}

Search for supersymmetric particles at colliders
usually proceeds from the assumption that all of them are relatively
heavy (few hundreds of GeV), depending on the values of soft supersymmetry
breaking mass parameters $m_0$, $m_{1/2}$, $A$ and short-living. Being heavier than the
Standard Model particles they usually decay faster and result in
usual particles with additional missing energy
taken away by the neutral stable lightest supersymmetric particle (LSP) - neutralino.
This is true almost in all the regions of
parameter space of the Minimal supersymmetric Standard Model (MSSM)
and for various mechanisms of supersymmetry
breaking~\cite{mssm1,mssm2,mssm3,mssm4}.

There exists, however, some regions in parameter space where the LSP
is not the usual neutralino, but a slepton (mainly stau) or the
relatively light superpartner of the $t$-quark (stop).
These regions are obviously considered as forbidden ones since the charged
LSP would be in conflict with astrophysical observations: no charged clouds of stable
particles are observed. At the border of these regions staus and stops become heavier than the neutralino
and thus unstable. Then they decay very fast.

Another constraint that is of great importance is the relic
density one. Given the amount of the dark matter from the WMAP
experiment~\cite{wmap1,wmap2} one is left with a narrow band of allowed
region which goes along the stau border line, then along the Higgs
limit line and then along the radiative electroweak symmetry breaking line.

There are two regions consistent with WMAP constraint that we discuss below. The first one is the
co-annihilation region where neutralinos and
staus are almost degenerate~\cite{falk,profumo}, and in the early Universe they would annihilate and
co-annihilate resulting in a proper amount of the dark matter. The second region appears only
for large negative trilinear scalar coupling $A_0$ and is distinguished by the
light stops.

We found out that in the narrow band at the border the forbidden regions staus or stops might be
rather stable with the lifetime long enough to go through the detector,
or produce secondary decay vertices inside the detector. Due to relatively small masses
(approximately within the range 150 $-$ 850 GeV)
the production cross-section of long-living staus at LHC may reach a few per cent of pb,
and stop production cross-sections can be as large as tens or even hundreds pb.

\section{MSSM co-annihilation region\\ and long-living staus}
\label{sec:stau}

\begin{figure}
\includegraphics[width=0.44\textwidth]{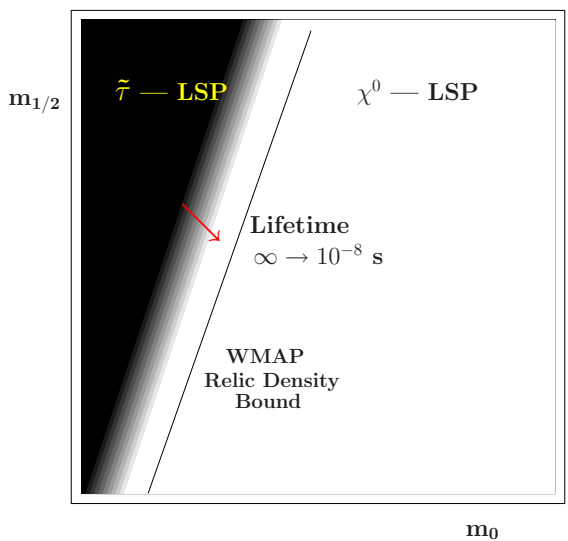}
\caption{LSP constraint in the $m_0 - m_{1/2}$ plane. Dark triangle
shows the region where stau is the LSP. At the boundary, the stau
lifetime decreases from left to right. The WMAP bound is shown as a
straight line.}
\label{fig:gladyshev_susy07_fig1}
\end{figure}

The co-annihilation region is shown qualitatively in the
$m_0 - m_{1/2}$ plane, Fig.~\ref{fig:gladyshev_susy07_fig1}. The dark triangle shows the region
where stau is the LSP. To the right of it the neutralino is the LSP.
The WMAP constraint goes along the LSP triangle border and is shown as a straight line.

\begin{figure}
\includegraphics[width=0.45\textwidth]{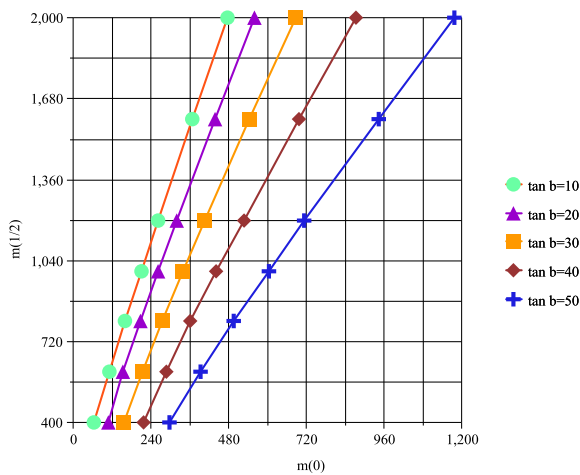}
\caption{$\tan\beta$ dependence of the LSP allowed region. To the left of the border stau
is the LSP and to the right -- neutralino is the LSP. The value of $\tan\beta$ increases
from left to right.}
\label{fig:gladyshev_susy07_fig2}
\end{figure}

\begin{figure}
\includegraphics[width=0.45\textwidth]{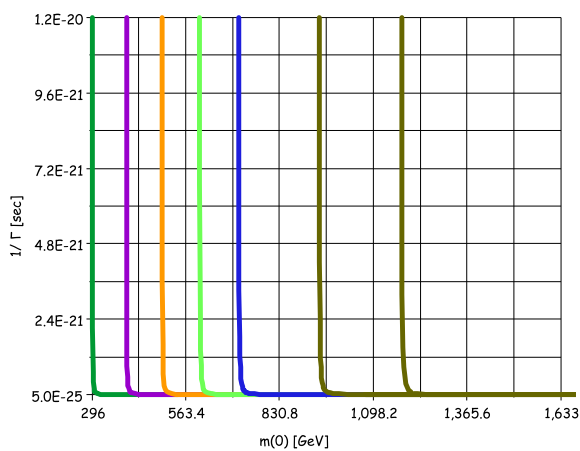}
\caption{The lifetime of stau as a function of $m_0$ near the border line for
$\tan\beta=50$. $m_{1/2}$ increases from left to right.}
\label{fig:gladyshev_susy07_fig3}
\end{figure}

Though the boundary of the LSP region with the WMAP allowed band is very narrow, its
position depends on the value of $\tan\beta$. In Fig.~\ref{fig:gladyshev_susy07_fig2} we show how the LSP
triangle increases  with  $\tan\beta$. Hence, even if it is very difficult to
get precisely into this narrow band, changing $\tan\beta$ one actually sweeps up a wide
area.

The boundary region happens to be a transition region from the stau-LSP to the neutralino-LSP.
In this very narrow zone the lifetime of stau rapidly changes from infinity to almost
zero passing the tiny interval (smeared by the change of $\tan\beta$) where stau is a
long-living particle.

When the stau mass becomes larger than that of the neutralino, stau decays. The only decay
mode in this region in case the $R$-parity is conserved is
$
\widetilde{\tau}\to \widetilde{\chi}^{0}_{1}\tau .
$
The life time crucially depends on the mass difference between
$\widetilde{\tau}$ and $\widetilde{\chi}^{0}_{1}$ and quickly
decreases while departing from the boundary line.
If we neglect mixing in the stau sector, then the next-to-lightest supersymmetric particle (NLSP) is the $\tilde\tau_1$
and the decay width is given by~\cite{bartl}
{\small $$
\mathrm\Gamma (\tilde\tau \! \to \! \chi^0_1 \tau) =
\frac{\alpha}{2} \left( N_{11}-N_{12}\tan\theta_W \right)^2 m_{\tilde\tau}
\left( 1-\frac{m_{\chi^0_1}^2}{m_{\tilde\tau}^2} \right)^{\!2} \!\! ,
$$}
where $N_{11}$ and $N_{12}$ are the elements of the matrix diagonalizing the
neutralino mass matrix.

In Fig.~\ref{fig:gladyshev_susy07_fig3} we show the lifetime of stau as a function of
$m_0$ for different values of $m_{1/2}$ and $\tan\beta=50$
calculated with the help of the ISAJET V7.67 code~\cite{isajet}.
One can see that a small deviation from the border line results
in immediate fall down of the lifetime. To get reasonable
lifetimes so that particles can
go through the detector or decay in the secondary vertices one needs to be almost exactly at
the borderline. However, the border line itself is not fixed,
it moves with $\tan\beta$~\cite{gkp1}.

\section{MSSM with Large Negative Values of $A$ and long-living stops}
\label{sec:stop}

Another interesting region of parameter space is the one distinguished by the
light stops. It appears only for large negative trilinear soft supersymmetry
breaking parameter $A_0$. On the border of this region, in full analogy with the
stau co-annihilation region, the top squark becomes the LSP and near
this border one might get the long-living stops.

Projected to the $m_0 - m_{1/2}$ plane the position of this region
depends on the values of $\tan\beta$ and $A$. In case when $|A|$ is
large enough the squarks of the third generation, and first of all
the lightest stop $\tilde t_1$, become relatively light. This happens via the see-saw
mechanism while diagonalizing the stop mass matrix
{\small
$$
\left(\begin{array}{cc} \tilde m_{t_L}^2& m_t(A_t-\mu\cot \beta )
\\ m_t(A_t-\mu\cot \beta ) & \tilde m_{t_R}^2 \end{array}  \right) \! ,
$$
}
where
{\small
\begin{eqnarray*}
  \tilde m_{t_L}^2&=&\tilde{m}_Q^2+m_t^2+\frac{1}{6}(4M_W^2-M_Z^2)\cos
  2\beta ,\\
  \tilde m_{tR}^2&=&\tilde{m}_U^2+m_t^2-\frac{2}{3}(M_W^2-M_Z^2)\cos
  2\beta .
\end{eqnarray*}
}
The off-diagonal terms increase with $A$, become large for large
$m_q$ (that is why the third generation) and give negative
contribution to the lightest top squark mass defined by minus sign in
{\small
$$
\tilde{m}^2_{1,2} \! = \! \frac12 \! \left( \! \tilde m_{t_L}^2 \!\!\! + \! \tilde m_{t_R}^2
\! \pm \! \sqrt{ \! \bigl( \! \tilde m_{t_L}^2 \!\!\! - \! \tilde m_{t_R}^2 \! \bigr)^{\!2} \!\!\!
+ \! 4m_t^2\bigl( \! A_t \! - \! \mu \!
\cot \! \beta \! \bigr)^{\!2}}\right) \!\! .
$$
}

\begin{figure*}
\begin{center}
\includegraphics[width=0.45\textwidth]{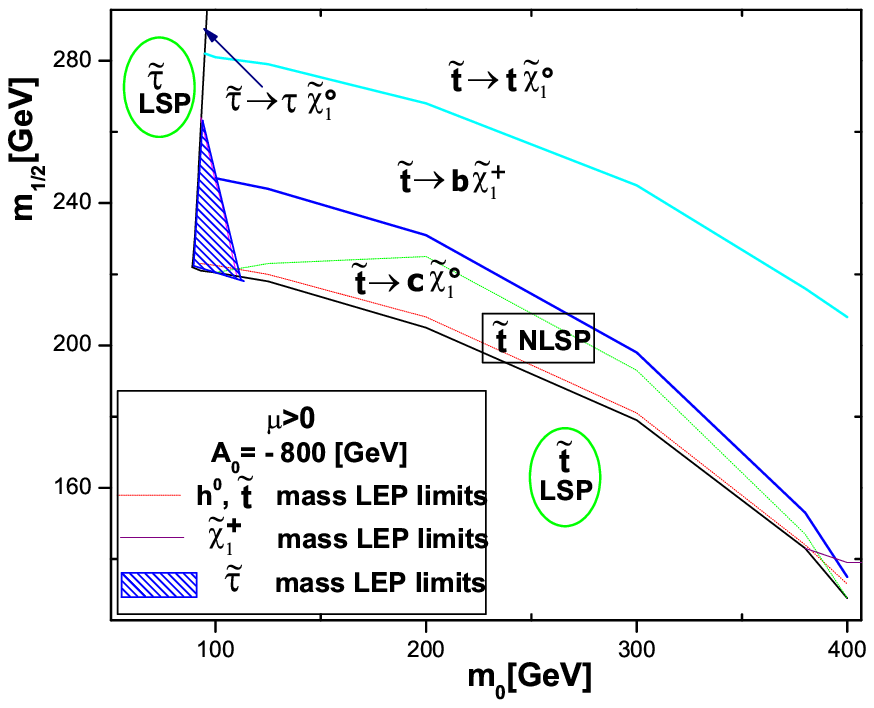}
\includegraphics[width=0.45\textwidth]{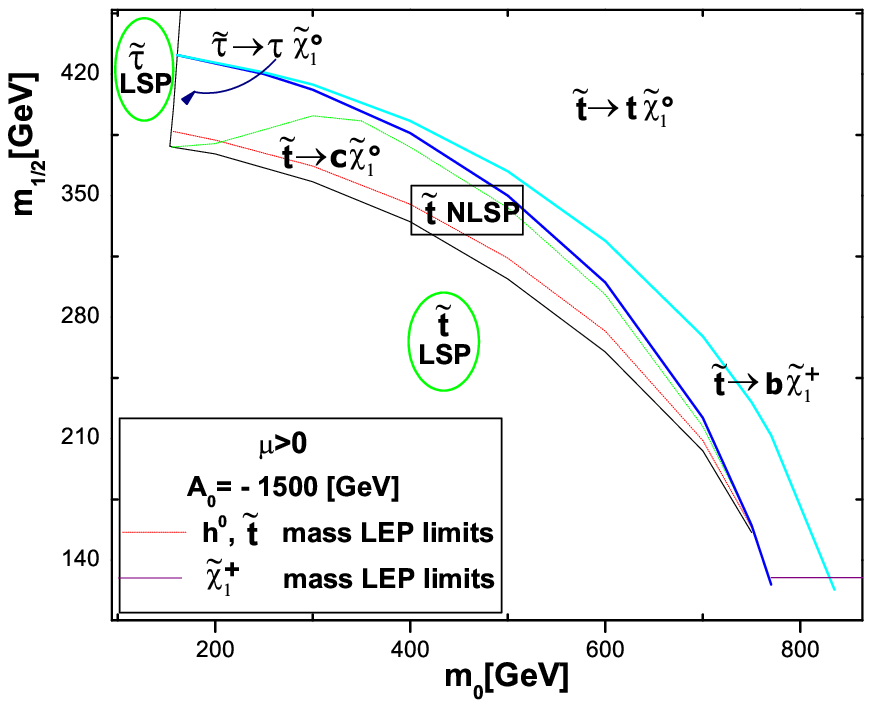}
\includegraphics[width=0.45\textwidth]{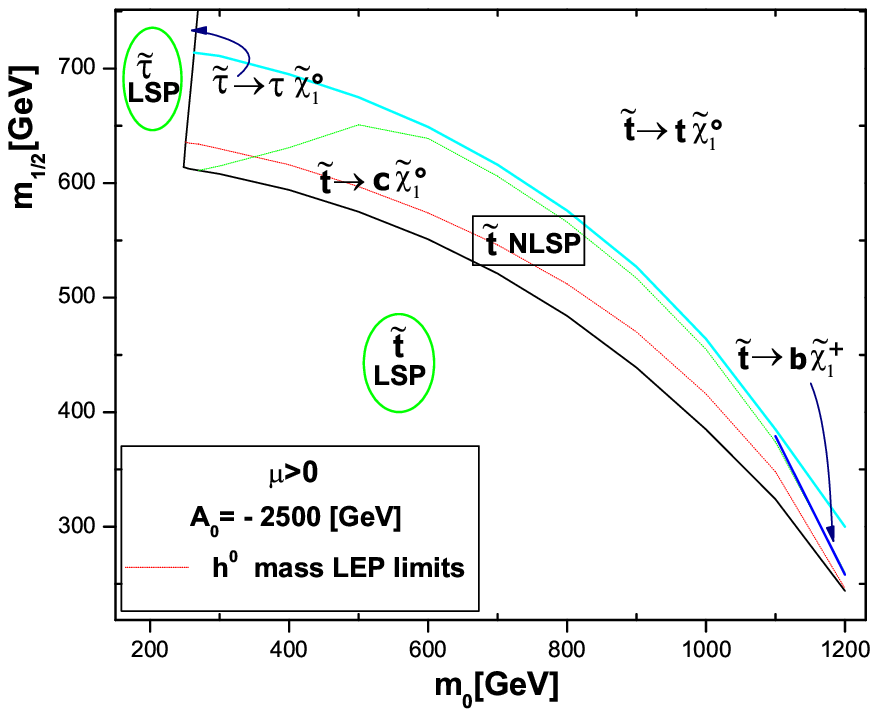}
\includegraphics[width=0.45\textwidth]{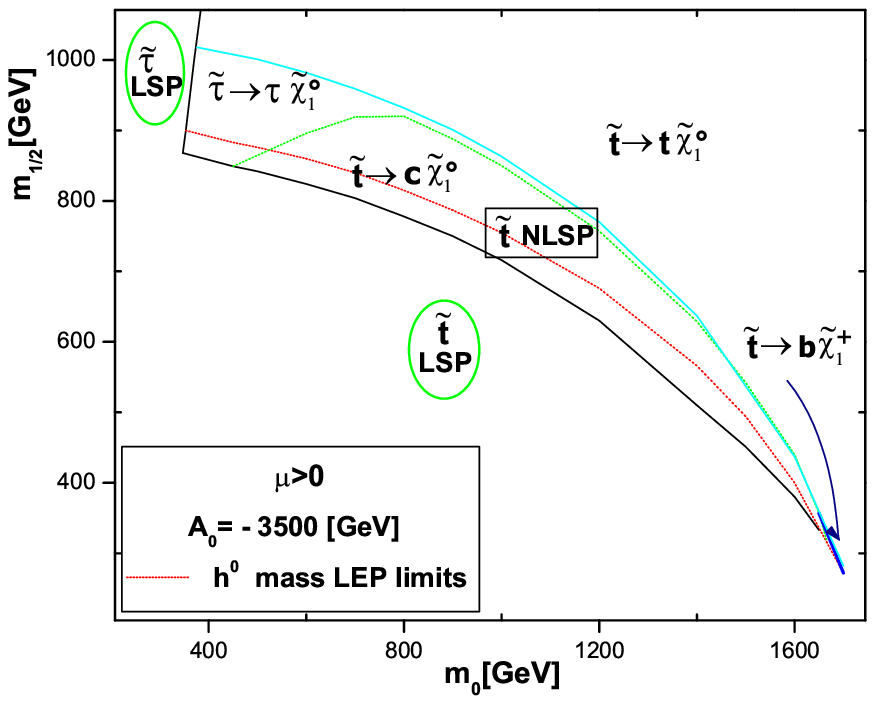}
\end{center}
\caption{Allowed region of the mSUGRA parameter space for
$A_0= -800, -1500, -2500, -3500$ GeV  and $\tan\beta=10$.  At the left from the
border stau is an LSP, below the border stop is the LSP. The dotted
line is the LEP Higgs mass limit. Also shown are the contours where
various stop decay modes emerge.}
\label{fig:gladyshev_susy07_fig4}       
\end{figure*}

Hence, increasing $|A|$ one can make the lightest stop as light as
one likes it to be, and even make it the LSP. The situation is
similar to that with stau for small $m_0$ and large $m_{1/2}$ when
stau becomes the LSP. For stop it takes place at small $m_0$
and small $m_{1/2}$. One actually gets the border line where stop becomes
the LSP. The region below this line is forbidden. It exists only
for large negative $A$, for small $A$ it is completely ruled out by
the LEP Higgs limit~\cite{gkp2}.

\begin{figure}
\includegraphics[width=0.44\textwidth]{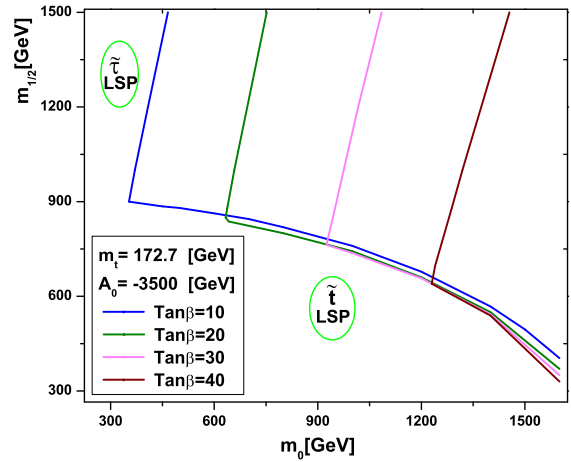}
\caption{Stau and stop constraints in the $m_0 - m_{1/2}$ plane for $A_0=-3500$ GeV
and different values of $\tan\beta$.}
\label{fig:gladyshev_susy07_fig5}       
\end{figure}

It should be noted that in this region  one gets not only the light
stop, but also the light Higgs, since the radiative correction to the Higgs
mass is proportional to the log of the stop mass.
The stop mass boundary is close to
the Higgs mass one and they may overlap for intermediate values of
$\tan\beta$. We show the projection of SUSY parameter space to the
$m_0 - m_{1/2}$ plane in Fig.~\ref{fig:gladyshev_susy07_fig4} for different values of $A$
and fixed $\tan\beta$. To calculate it we again used the ISAJET code~\cite{isajet}.

One can see that when $|A|$ decreases the border line moves down and
finally disappears. On the contrary, increasing $|A|$ one gets
larger forbidden area and the value of the stop mass at the border
increases.

Changing $\tan\beta$ one does not influence the stop border line,
the only effect is the shift of the stau border line which moves to the
right with increase of $\tan\beta$ as shown in Fig.~\ref{fig:gladyshev_susy07_fig5}, so
that the whole forbidden area increases and covers the left bottom
corner of the $m_0 - m_{1/2}$ plane.

It should be mentioned that the region near the border line is very
sensitive to the Standard Model parameters, a minor shift in $\alpha_s$ or $m_t$
and $m_b$ leads to noticeable change of spectrum as can be seen, for example, from
comparison of different codes~\cite{kraml}.

Calculation of the required relic density with the help of MicrOmegas
package~\cite{micromegas1,micromegas2} shows that again it is very sensitive
to the input parameters, however, since the stop border line is very
close to the Higgs limit, the relic density constraint may be met here
fitting $A_0$ and/or $\tan\beta$.

\section{Long-living superparticles at the LHC}
\label{sec:pheno}

Searches for long-living particles were already made by LEP
collaborations~\cite{searches1,searches2,searches3}.
It is also of great interest
at the moment since the first physics results of the coming LHC are
expected in the nearest future. Light long-living staus and stops could be produced already
during first months of its operation~\cite{stopprod1,stopprod2}.

Since staus and stops are relatively light in our scenario, the production cross-sections
are quite large and may achieve a few per cent of pb for stau production, and tens
or even hundreds of pb for light stops, $m_{\tilde t}< 150$~GeV.
The cross-section then quickly falls down when the mass of
stop is increased. However, even for very large values of $|A|$ when stops become heavier than several hundreds GeV,
the production cross-section is of order of few per cent of pb, which is enough for detection
with the high LHC luminosity.

Being created staus and stops decay.
As it was already mentioned, the only possible decay mode of stau is
$
\widetilde{\tau}\to \widetilde{\chi}^{0}_{1}\tau .
$
The top-squark has several different decay modes depending on its mass.
If stop is heavy enough it decays to the bottom quark and the lightest chargino
($\tilde t \to b \tilde\chi^\pm_1$). However, for large values of $|A|$,
namely $A_0<-1500$~GeV the region where this decay takes place is getting smaller
and even disappear due to mass inequality $m_{\tilde t} < m_b + m_{\tilde\chi_1^\pm}$
(see right bottom corner in Fig.~\ref{fig:gladyshev_susy07_fig4}). In this case the dominant decay mode is
the decay to the top quark and the lightest neutralino ($\tilde t \to t \tilde\chi^0_1$).
Light stop decays to the charm quark and the lightest neutralino
($\tilde t \to c \tilde\chi^0_1$)~\cite{stopdecay}. The latter decay, though it is loop-suppressed,
has the branching ratio 100 \%.

\section{Conclusions and discussions}
\label{sec:concl}

We have shown that within the framework of the Constrained MSSM with
gravity mediated soft supersymmetry breaking mechanism there exists
an interesting possibility to get long-living next-to-lightest supersymmetric
particles (staus and stops). Their production cross-sections crucially depend
on a single parameter, the mass of the superparticle, and for light staus can
reach a few \% pb. This might be within the LHC reach.
The stop production cross-section can achieve even hundreds pb.
Light stop NLSP scenario requires large negative values of the soft
trilinear SUSY breaking parameter $A$. Decays of long-living staus and stops would have an unusual
signature if heavy charged particles decay with a considerable delay in
secondary vertices inside the detector, or even escape the detector.
Stops can also form so-called $R$-hadrons (bound states of SUSY particles)
if their lifetime is bigger than the hadronisation time.

Experimental Higgs and chargino mass limits as well as WMAP relic density limit
can be easily satisfied in our scenario. However, the strong fine-tuning is required.
Stau-NLSP and stop-NLSP scenarios differ from the GMSB scenario~\cite{gmsb1,gmsb2}
with the gravitino as the LSP, and next-to-lightest supersymmetric particles typically live longer.

Further details and numerical results as well as complete list of references can be found in~\cite{gkp1,gkp2}.

\bigskip
Financial support from Russian Foundation for Basic Research (grants \# 05-02-17603, \# 08-02-00856)
and the Ministry of Education and Science of the Russian
Federation (grant \# 5362.2006.2) is kindly acknowledged.

\end{document}